\documentclass[12pt,a4paper]{article}

\newcommand{\pa}{\partial}
\newcommand{\te}{\theta}
\newcommand{\la}{\lambda}

\begin{document}

\begin{flushright}
{ }
\end{flushright}
\vspace{1.8cm}

\begin{center}
 \textbf{\Large The SU(1$|$2) and SU(2$|$2) Sectors \\
 from Superstrings in $AdS_5 \times S^5$}
\end{center}
\vspace{1.6cm}
\begin{center}
 Shijong Ryang
\end{center}

\begin{center}
\textit{Department of Physics \\ Kyoto Prefectural University of Medicine
\\ Taishogun, Kyoto 603-8334 Japan}
\par
\texttt{ryang@koto.kpu-m.ac.jp}
\end{center}
\vspace{2.8cm}
\begin{abstract}
For the $\kappa$-symmetry gauge fixed superstring action in 
$AdS_5 \times S^5$ we consider the fermionic fluctuations over
a circular bosonic string background with two angular 
momenta and two winding numbers in $S^5$. The SU(2)-type 
redefinitions of fermionic fields and the first-string limit
generate a truncated fermionic action for the SU(1$|$2)
sector. It is expressed in a two-dimensional Lorentz-invariant
form of a massive Dirac fermion and the plane-wave spectrum for
the fermionic excitations is derived. The fermionic spectrum
for the SU(2$|$2) sector is also analyzed.
\end{abstract}
\vspace{3cm}
\begin{flushleft}
April, 2006 
\end{flushleft}

\newpage
\section{Introduction}

The AdS/CFT correspondence \cite{MGW} has been explored beyond the 
supergravity approximation. Inspired from the solvability of the string
theory in the pp-wave background of $AdS_5 \times S^5$ \cite{MT},
it has been proposed that the energies of specific free massive string
excited states can be matched with the perturbative scaling dimensions of
gauge invariant near-BPS operators with large R-charge in the BMN limit
for the $\mathcal{N}=4$ SU(N) super Yang-Mills (SYM) theory \cite{BMN}.
The BMN result has been reproduced by the semiclassical quantization 
of nearly point-like string with large angular momentum along a central
circle in $S^5$ \cite{GKP}

The energies of various semiclassical extended string configurations
with several large angular momenta in $AdS_5 \times S^5$  have
been shown in \cite{FT,BMS,ART,AT} to match with the anomalous
dimensions of the corresponding long SYM non-BPS operators,
which can be computed by using the Bethe ansatz \cite{MZ} for
diagonalization of the dilatation operator \cite{NBS,NB,BZ},
that is represented by a Hamiltonian of an integrable spin chain. 

From the view point of integrability the gauge/string duality has 
been further confirmed by verifying the equivalence between the classical
string Bethe equation for the classical $AdS_5 \times S^5$ string
sigma model and the Bethe equation for the spin chain \cite{KMM,KZ}.
Combining the classical string Bethe ansatz and the all-loop
gauge theory asymptotic Bethe ansatz \cite{BDS}, it has been shown
that a novel Bethe ansaz, namely, the quantum string Bethe ansatz for the
SU(2) sector has been constructed \cite{AFS} such that it generates the
classical spinning strings, the $\lambda^{1/4}$ strong coupling 
asymptotics and the $1/J$ energy corrections of arbitrary $M$-impurity
BMN states whose special $M = 2, 3$ cases agree with the results
of direct light-cone gauge quantization of the interacting string theory
in the near plane-wave background \cite{CLM}.
The gauge theory asymptotic Bethe ansatz for the SU(2) sector has been
shown to arise as an approximation to the Hubbard model \cite{RSS}.
The quantum string Bethe ansatz has been generalized by constructing the
S matrices to  the other sectors such as SL(2), SU(1$|$1) \cite{MS}
and the full PSU(2,2$|$4) \cite{BS}. 
For the SU(1$|$1) sector the dilatation
operator at one-loop has been shown to coincide with the Hamiltonian
of the free lattice fermion \cite{CHM}. 
The S matrices leading to the
asymptotic Bethe equations have been investigated for the SU(1$|$2) and
SU(2$|$2) sectors \cite{NBT} and the two-loop dilatation operator for
the SU(1,1$|$2) sector has been constructed \cite{BIZ}.
 
On the other hand there have been various studies of comparing the quantum
world-sheet corrections to spinning string solutions 
in $AdS_5 \times S^5$ with the finite size corrections to the Bethe 
equations \cite{FPT,SZ}. 

The gauge/string duality has been also presented at the level of equations
of motion \cite{AM} and at the level of effective action \cite{MK} where
an interpolating spin chain sigma model action constructed by taking
the continuum limit of the spin chain in the coherent basis for
the SU(2) sector is also reproduced by taking some fast-string limit of
the string action. The latter approach has been extended to the whole 
SO(6) and its compact subgroups \cite{KT,KRT,ST} and non-compact 
SL(2) \cite{ST,BM}. Based on the spin chain sigma model for the SU(2) 
sector the $1/J$ and $1/J^2$ energy corrections to  the plane-wave state
and the circular and folded string states have been computed \cite{MTT}.
The supersymmetric extensions have been performed for SU(1$|$3) \cite{HL},
SU(1,1$|$1) \cite{BCM}, SU(1,1$|$2) \cite{BC} and SU(2$|$3) \cite{HL,BST}.
In \cite{HL,BCM,BC} the first-string limit has been taken for certain
subsectors of the covariant $\kappa$-symmetric superstring action
in $AdS_5 \times S^5$ \cite{RMT} 
constructed as a 2d sigma model on the coset
superspace PSU(2,2$|$4)/[SO(1,4)$\times$SO(5)], while in \cite{BST}
it has been taken for a subsector of the light-cone $\kappa$-symmetry
gauge fixed superstring action \cite{MAT} expressed in terms of the
light-cone supercoset coordinates in the SU(3)$\times$U(1) invariant 
form. 

From the superstring sigma model action in $AdS_5 \times S^5$ expressed
in terms of the $Z_4$-graded current of 
the PSU(2,2$|$4)/[SO(1,4)$\times$SO(5)]
supercoset \cite{RS}, the truncations to the SU(1$|$1) sector
have been performed by choosing a phase-space uniform gauge $t=\tau,
p_{\phi} =J$ \cite{AAF} where $p_{\phi}$ is the canonical momentum
conjugate to the angle variable $\phi$ for a central circle in $S^5$
and a uniform light-cone gauge \cite{GAF}, where the BMN spectrum
for fermions is presented and the $1/J$ correction to the $M$-impurity
plane-wave state agrees with the result of \cite{TMS}.
For the former gauge choice the two complex fermions are arranged into 
a single world-sheet Dirac fermion so that the reduced action shows
a non-trivial 2d Lorentz-invariant interacting theory of massive
Dirac fermion, while for the latter gauge choice the reduced theory
becomes free and the femionic fluctuation spectra over both a 
point-like string with no winding numbers and an extended string wound
around the $\phi$ direction are computed.
For the former truncation to the SU(1$|$1) sector the exact S-matrix
has been computed to give the Bethe ansatz solution \cite{KKZ}.

In ref. \cite{BST} the SU(1$|$1) sector has been extracted from the 
SU(3)$\times$U(1) invariant superstring action, where the fermionic action
for the quadratic fermionic fluctuation over the point-like
bosonic string background with a large angular momentum becomes
a non-relativistic expansion form of an action for a massive
2d relativistic fermion. In order to see the effect of the bosonic
background on the fermionic fluctuation we will consider the SU(1$|$2) 
sector. We will study the fermionic fluctuation around the extended 
circular string background specified by two angular momenta and two
winding numbers in $S^5$, and take the large limit of the total angular
momentum. The fermionic action simplified by taking the first-string 
limit will be expressed in an explicitly 2d relativistic manner by a
massive world-sheet Dirac fermion and shown 
to have the plane-wave spectrum.
The fermionic spectrum for the SU(2$|$2) sector will be discussed.

\section{SU(3)$\times$U(1) invariant superstring action}

We consider the superstring in $AdS_5 \times S^5$ space-time with metric
$ds^2 = e^{2\phi}dx^adx^a + d\phi d\phi + dX^MdX^M, \;
X^MX^M = 1 \; (a=0, \cdots,3; M=1, \cdots, 6)$. 
In terms of the light-cone coordinates on the
PSU(2,2$|$4)/[SO(1,4)$\times$SO(5)] supercoset the full Lagrangian
$\mathcal{L} = \mathcal{L}_{kin} + \mathcal{L}_{WZ}$ in the fermionic
light-cone $\kappa$-symmetry gauge is constructed as \cite{MAT}
\begin{eqnarray}
\mathcal{L}_{kin} &=& -\frac{1}{2}\sqrt{g} g^{\mu\nu}[2e^{2\phi}
(\pa_{\mu}x^+\pa_{\nu}x^- + \pa_{\mu}x\pa_{\nu}\bar{x}) + 
\pa_{\mu}\phi\pa_{\nu}\phi + \pa_{\mu}X^M\pa_{\nu}X^M ] \nonumber \\
&-& \frac{i}{2}\sqrt{g} g^{\mu\nu}e^{2\phi}\pa_{\mu}x^+ [
\te^A\pa_{\nu}\te_A + \te_A\pa_{\nu}\te^A +  \eta^A\pa_{\nu}\eta_A + 
\eta_A\pa_{\nu}\eta^A ] \nonumber \\
&-& i\sqrt{g} g^{\mu\nu}e^{2\phi}\pa_{\mu}x^+X^N\pa_{\nu}X^M\eta_A
{\rho^{MNA}}_B\eta^B \nonumber \\
&+& \frac{1}{2}\sqrt{g} g^{\mu\nu}e^{4\phi}\pa_{\mu}x^+\pa_{\nu}x^+[
(\eta^A\eta_A)^2 + (X^N\eta_A{\rho^{MNA}}_B\eta^B)^2 ], \nonumber \\
\mathcal{L}_{WZ} &=& \epsilon^{\mu\nu} e^{2\phi}\pa_{\mu}x^+X^M
(\eta^A\rho^M_{AB}\pa_{\nu}\te^B + \eta_A\rho^{MAB}\pa_{\nu}\te_B)
 \nonumber \\
&+& i\sqrt{2}\epsilon^{\mu\nu} e^{3\phi}\pa_{\mu}x^+X^M(\pa_{\nu}\bar{x}
\eta_A\rho^{MAB}\eta_B - \pa_{\nu}x\eta^A\rho^M_{AB}\eta^B),
\label{lag}\end{eqnarray}
where $g_{\mu\nu} \;(\mu = 0,1)$ is a world-sheet metric with signature 
$(-,+)$ and $g =- \det g_{\mu\nu}$.  The Poincare coordinates of $AdS_5$
are chosen by 
\begin{equation}
x^{\pm} = \frac{1}{\sqrt{2}}(x^3 \pm x^0), \hspace{1cm}  x =
\frac{1}{\sqrt{2}}(x^1 + ix^2), \hspace{1cm} \bar{x} =
\frac{1}{\sqrt{2}}(x^1 - ix^2) 
\end{equation}
and the radial direction $\phi$, while $S^5$ is parametrized by a unit 
6-vector $X^M$ so that the constraint 
$X^MX^M = 1$ should be imposed with a
Lagrange multiplier $\Lambda$. This Lagrangian has manifest SU(4) symmetry
where the $4+ 4$ complex fermionic fields $\te_A, \eta_A$ with $A=1,2,3,4$
transform in the fundamental representation of SU(4) and
$\te^A = \te_A^{\dagger}, \eta^A = \eta_A^{\dagger}$. The 
$4\times 4$ matrices $\rho^M$ are
``off-diagonal" blocks of the SO(6) gamma-matrices in the chiral 
representation and $\rho^{MN} = - \rho^{[M}\rho^{*N]}$.
There exist only quadratic and quartic fermionic terms which are 
associated with special symmetries of the $AdS_5 \times S^5$
background. The $(\eta^A\eta_A)^2$ term reflects the curvature of the
background, and the $(X^N\eta_A{\rho^{MNA}}_B\eta^B)^2$ term is
interpreted as the coupling to the R-R 5-form background. The fermionic
fields $\te_A$ are related to the linearly realized supersymmetry of
the super conformal algebra PSU(2,2$|$4), and the $\eta_A$ fields are 
associated with the non-linearly realized superconformal 
symmetry. 

In ref. \cite{BST} the bosonic $AdS_5$ Poincare coordinates are replaced
by the global $AdS_5$ ones and in order to choose the conformal gauge
for the 2d metric the following ansatz corresponding to the global
$AdS_5$ time $t = \kappa \tau + \cdots$ where dots indicate possible
fermionic terms, is taken for the bosonic $AdS_5$ fields 
\begin{equation}
e^{\phi} = \cos \kappa \tau, \hspace{0.5cm} x^+ = \frac{\tan\kappa\tau}
{\sqrt{2}}, \hspace{0.5cm}  x^- = -\frac{\tan\kappa\tau}{\sqrt{2}} + 
f(\tau,\sigma), \hspace{0.5cm} x=\bar{x} = 0. 
\label{ans}\end{equation}
The $\phi$ equation of motion restricts on the allowed fermionic 
configuration to determine $\pa_0f$, while one of the two conformal
constraints determines $\pa_1f$.

In the SU(2$|$3) sector on the SYM side the gauge invariant operators
consist of the 3 chiral complex combinations of 6 scalars on which
the SO(6) R-symmetry acts and the two spinor components of the gluino
Weyl fermion which are singlets under the Cartan $[U(1)]^3$ subgroup of
SO(6). In order to extract the SU(2$|$3) sector the 3 chiral bosonic
fields $X_i$ are introduced by $X_i \equiv X_{2i-1} + iX_{2i}, \; i=1,2,3$
and the SU(4) fermions are splitted in $3+1$ way as 
$\eta_A \equiv (\eta_i, \eta), \; \te_A \equiv (\te_i, \te), i=1,2,3$.
The two SU(3) singlet fields $\eta \equiv \eta_4, \te \equiv \te_4$
are related with the two fermions in the SU(2$|$3) sector.

The fermionic part of the Lagrangian (\ref{lag}) can be expressed in the
manifestly SU(3)$\times$U(1) invariant form through the ansatz 
(\ref{ans}) as $\mathcal{L}_F = \mathcal{L}_{2F} + \mathcal{L}_{4F}$
where the quadratic terms are 
\begin{eqnarray}
\mathcal{L}_{2F} &=& \frac{\kappa}{\sqrt{2}}[ i\eta^i\pa_0\eta_i + 
i\bar{\eta}\pa_0\eta + i\te^i\pa_0\te_i + i\bar{\te}\pa_0\te
+ \epsilon_{ijk}\eta^i\pa_1\te^jX^k - \epsilon^{ijk}\eta_i\pa_1\te_jX_k 
\nonumber \\   
&+& \eta^i\pa_1\bar{\te}X_i - \eta_i\pa_1\te X^i + \pa_1\te^i\bar{\eta}X_i
- \pa_1\te_i\eta X^i \nonumber \\
&-& i(X^i\pa_0X_j - X_j\pa_0X^i)\eta_i\eta^j - iX^i\pa_0X_i(\eta^j\eta_j
- \bar{\eta}\eta ) \nonumber \\
&-& i(\epsilon^{ijk}X_j\pa_0X_k\eta_i \bar{\eta} - 
\epsilon_{ijk}X^j\pa_0X^k\eta \eta^i) ]
\label{twl}\end{eqnarray}
and the quartic terms are 
\begin{eqnarray}
\mathcal{L}_{4F} &=& -\left(\frac{\kappa}{\sqrt{2}}\right)^2 [ 
3\eta^i\eta_i\bar{\eta}\eta - 4 X_i\eta^iX^j\eta_j\bar{\eta}\eta
+ 4 \eta_iX^i\eta^jX_j\eta_k\eta^k \nonumber \\
&+& 2\epsilon_{ijk}\eta^i\eta^jX^k\eta_lX^l\eta + 2\epsilon^{ijk}
\eta_i\eta_jX_k\eta^lX_l\bar{\eta} ],
\label{fol}\end{eqnarray}
where $X^i = X_i^*, \eta^i = \eta_i^{\dagger}, \te^i = \te_i^{\dagger},
\bar{\eta}=\eta^{\dagger}, \bar{\te}=\te^{\dagger}$ and there 
are no gamma-matrices.
The coupling terms including the $\sigma$-derivative in (\ref{twl}) 
originate in the Wess-Zumino part $\mathcal{L}_{WZ}$ of (\ref{lag}),
while the coupling terms including the $\tau$-derivative in the form
$X\pa_0 X\eta \eta$ are due to the quadratic terms proportional to
$\eta_A{\rho^{MNA}}_B\eta^B$ in the kinetic part $\mathcal{L}_{kin}$ of
(\ref{lag}). The bosonic part of the Lagrangian (\ref{lag}) is also
expressed as
\begin{equation}
\mathcal{L}_B = -\frac{1}{2}\pa^{\mu}X_i^*\pa_{\mu}X_i +
\frac{1}{2} \Lambda (X_i^*X_i - 1).
\label{bol}\end{equation}

\section{Fermionic fluctuations over a circular string with two
equal spins}

There are two possible consistent truncations A and B 
with the bosonic fields from
$AdS_3 \times S^3$ and fermions suitably chosen as
\begin{equation}
\mathrm{A}:\; (X_1, X_2; \te, \te_3, \eta_1, \eta_2) \ne 0, \hspace{1cm}
(x, X_3; \eta, \eta_3, \te_1, \te_2) = 0,
\end{equation}
or, alternatively,
\begin{equation}
\mathrm{B}:\; (X_1, X_2; \eta, \eta_3, \te_1, \te_2) \ne 0, \hspace{1cm}
(x, X_3; \te, \te_3, \eta_1, \eta_2) = 0.
\end{equation}
This ``SU(1$|$2)" string theory sector is considered to be related with
the SU(1$|$2) gauge theory sector. Further restricting to $S^1$ inside
$S^5$ we have the following two truncations that are associated with
the SU(1$|$1) gauge theory sector
\begin{eqnarray}
\mathrm{A'} &:& \hspace{1cm} (X_1; \te, \eta_1) \ne 0, \\
\mathrm{B'} &:& \hspace{1cm} (X_1; \eta, \te_1) \ne 0,
\end{eqnarray}
where the other bosonic and fermionic 
fields are switched off respectively.
Some solitonic classical solutions for these subsectors were constructed
to include the fermionic  semiclassical contribution 
as the generalization of the bosonic spinning string solutions 
\cite{BST}. For the ``SU(1$|$1)" string theory sectors A' and B'
the  non-relativistic actions of BMN-type massive fermionic
fluctuations were constructed from the SU(3)$\times$U(1) invariant
Lagrangian (\ref{twl}), (\ref{fol}) and (\ref{bol}) in the point-like
bosonic string background  by integrating extra fermions $\eta_1$ and
$\te_1$ respectively.

As an extended string background we prepare a circular string solution
specified by the winding number $m$ with large two equal spins 
$\mathcal{T}_1 = \mathcal{T}_2 = \mathcal{T}/2$, which is expressed as
\begin{equation}
X_1 = \frac{1}{\sqrt{2}}e^{i\omega\tau - im\sigma}, \hspace{1cm}
X_2 = \frac{1}{\sqrt{2}}e^{i\omega\tau + im\sigma}
\label{bob}\end{equation}
with $\omega = \mathcal{T}$ \cite{ART}. Its energy is characterized by
$\mathcal{E}^2 = \kappa^2 = \mathcal{T}^2 + m^2$. This string background
is determined from the leading order relations of the conformal 
constraints where there are no fermionic semiclassical contributions. 
We consider the fermionic excitation representing 
a small perturbation over the circular bosonic string background.
This configuration is mapped to the long SYM operator which is composed of
the large and same number of two comlex scalar fields and only a few
fermions. It is convenient to rescale the fermionic fields as
\begin{equation}
\left(\begin{array}{c} \eta_i \\ \eta^i \end{array} \right) \rightarrow
\alpha \left(\begin{array}{c} \eta_i \\ \eta^i \end{array} \right), \;
\left(\begin{array}{c} \eta \\ \bar{\eta} \end{array} \right) \rightarrow
\alpha\left(\begin{array}{c} \eta \\ \bar{\eta} \end{array} \right), \;
\left(\begin{array}{c} \te_i \\ \te^i \end{array} \right) \rightarrow
\alpha\left(\begin{array}{c} \te_i \\ \te^i \end{array} \right), \;
\left(\begin{array}{c} \te \\ \bar{\te} \end{array} \right)
\rightarrow \alpha \left(\begin{array}{c} \te \\ \bar{\te} 
\end{array} \right)
\end{equation}
with $\alpha = (\sqrt{2}/\kappa)^{1/2}$ in order 
to absorb the overall factors $\kappa/\sqrt{2}$ in $\mathcal{L}_{2F}$
and $(\kappa/\sqrt{2})^2$ in $\mathcal{L}_{4F}$. 

Here we start to consider the case A.
Plugging the bosonic background (\ref{bob}) into the quadratic Lagrangian
(\ref{twl}) with $\kappa/\sqrt{2}=1$ we have
\begin{eqnarray}
\mathcal{L}_{2F} &=& i\sum_{i=1}^2 \eta^i\pa_0\eta_i + i\te^3\pa_0\te_3
+ i\bar{\te}\pa_0\te  \nonumber \\
&-& \frac{e^{-i\omega\tau}}{\sqrt{2}}(e^{-im\sigma}\eta^1 - 
e^{im\sigma}\eta^2)\pa_1\te^3 + \frac{e^{i\omega\tau}}{\sqrt{2}}
(e^{im\sigma}\eta_1 - e^{-im\sigma}\eta_2)\pa_1\te_3 \nonumber \\
&+& \frac{e^{i\omega\tau}}{\sqrt{2}}(e^{-im\sigma}\eta^1 + 
e^{im\sigma}\eta^2)\pa_1\bar{\te} - \frac{e^{-i\omega\tau}}{\sqrt{2}}
(e^{im\sigma}\eta_1 + e^{-im\sigma}\eta_2)\pa_1\te \nonumber \\ 
&+& \omega(e^{2im\sigma}\eta_1\eta^2 + e^{-2im\sigma}\eta_2\eta^1).
\label{lom}\end{eqnarray}
The quartic Lagrangian (\ref{fol}) becomes
\begin{equation}
\mathcal{L}_{4F} = -4 \eta_1\eta^1\eta_2\eta^2.
\label{let}\end{equation}
In view of the terms including the $\sigma$-derivative in (\ref{lom})
which are associated with the Wess-Zumino part we introduce a change
of fermionic variables
\begin{equation}
\left(\begin{array}{c} \eta_- \\ \eta_+ \end{array} \right) =
\frac{1}{\sqrt{2}} \left(\begin{array}{cc} e^{im\sigma} & -e^{-im\sigma}
\\ e^{im\sigma} & e^{-im\sigma} \end{array} \right)
\left(\begin{array}{c} \eta_1 \\ \eta_2 \end{array} \right), 
\label{cha}\end{equation}
which is an SU(2)-type rotation. We use $\omega = \mathcal{T}$ and
the inverse relation of (\ref{cha}) to rewrite (\ref{lom}) as
\begin{eqnarray}
\mathcal{L}_{2F} &=& i\eta^+\pa_0\eta_+ + i\bar{\te}\pa_0\te + 
e^{i\mathcal{T}\tau}\eta^+\pa_1\bar{\te} - e^{-i\mathcal{T}\tau}
\eta_+\pa_1\te - \mathcal{T}\eta^+\eta_+ \nonumber \\
&+& i\eta^-\pa_0\eta_- + i\te^3\pa_0\te_3 - 
e^{-i\mathcal{T}\tau}\eta^-\pa_1\te^3 + e^{i\mathcal{T}\tau}
\eta_-\pa_1\te_3 + \mathcal{T}\eta^-\eta_-,
\label{lpm}\end{eqnarray}
where $\eta^+ = \eta_+^{\dagger}, \eta^- = \eta_-^{\dagger}$ and
the $\sigma$-dependent exponential phase factors $e^{\pm 2im\sigma}$
in the mixed terms in (\ref{lom}) have been eliminated. 
Under the SU(2)-type rotation (\ref{cha}) the kinetic terms remain
the canonical forms. In the expression $\mathcal{L}_{2F}$
(\ref{lom}) the fermionic field $\eta_1$ is coupled with 
$\eta^2$, and $\eta_2$ with 
$\eta^1$, while the transformed one (\ref{lpm}) is simplified such that
$\eta_+$ is separated from $\eta_-$. The quartic expression (\ref{let})
is changed into
\begin{equation}
\mathcal{L}_{4F} = -4 \eta^+\eta_+\eta^-\eta_-.
\label{lfp}\end{equation}
Making the scalings of $\te$ and $\te_3$ as $\te \rightarrow 
e^{i\mathcal{T}\tau}\te, \te_3 \rightarrow 
e^{-i\mathcal{T}\tau}\te_3$ to remove the time-dependent exponential phase
factors in the mixed terms, we have a symmetric expression
\begin{eqnarray}
\mathcal{L}_{2F} &=& i\eta^+\pa_0\eta_+ + i\bar{\te}\pa_0\te + 
\eta^+\pa_1\bar{\te} - \eta_+\pa_1\te - \mathcal{T}(\bar{\te}\te + 
\eta^+\eta_+)  \nonumber \\
&+& i\eta^-\pa_0\eta_- + i\te^3\pa_0\te_3 - 
(\eta^-\pa_1\te^3 - \eta_-\pa_1\te_3) + \mathcal{T}(\te^3\te_3 + 
\eta^-\eta_-).
\label{syl}\end{eqnarray}

Let us turn to the case B. The quadratic Lagrangian (\ref{twl}) in the 
bosonic background (\ref{bob}) is written by
\begin{eqnarray}
\mathcal{L}_{2F} &=& i\sum_{i=1}^2 \te^i\pa_0\te_i + i\eta^3\pa_0\eta_3
+ i\bar{\eta}\pa_0\eta  \nonumber \\
&+& \frac{e^{-i\omega\tau}}{\sqrt{2}}\eta^3(e^{-im\sigma}\pa_1\te^1 - 
e^{im\sigma}\pa_1\te^2) - \frac{e^{i\omega\tau}}{\sqrt{2}}\eta_3
(e^{im\sigma}\pa_1\te_1 - e^{-im\sigma}\pa_1\te_2) \nonumber \\
&+& \frac{e^{i\omega\tau}}{\sqrt{2}}(e^{-im\sigma}\pa_1\te^1 + 
e^{im\sigma}\pa_1\te^2)\bar{\eta} - \frac{e^{-i\omega\tau}}{\sqrt{2}}
(e^{im\sigma}\pa_1\te_1 + e^{-im\sigma}\pa_1\te_2)\eta \nonumber \\ 
&+& \omega(\eta^3\eta_3 - \bar{\eta}\eta),
\label{ltb}\end{eqnarray}
while the quartic one (\ref{fol}) is
\begin{equation}
\mathcal{L}_{4F} = -3\eta^3\eta_3\bar{\eta}\eta.
\label{lft}\end{equation}
The expression (\ref{ltb}) also suggests the following SU(2)-type rotation
\begin{equation}
\left(\begin{array}{c} \te_- \\ \te_+ \end{array} \right) =
\frac{1}{\sqrt{2}} \left(\begin{array}{cc} e^{im\sigma} & -e^{-im\sigma} 
\\ e^{im\sigma} & e^{-im\sigma} \end{array} \right)
\left(\begin{array}{c} \te_1 \\ \te_2 \end{array} \right), 
\label{chb}\end{equation}
which corresponds to (\ref{cha}). This change of variables leads to an
expression with no $\sigma$-dependent exponential factors
\begin{eqnarray}
\mathcal{L}_{2F} &=& i\te^+\pa_0\te_+ + i\bar{\eta}\pa_0\eta
+ e^{i\mathcal{T}\tau}\pa_1\te^+\bar{\eta}  - e^{-i\mathcal{T}\tau}
\pa_1\te_+ \eta - \mathcal{T}\bar{\eta}\eta \nonumber \\
&+& i\te^-\pa_0\te_- + i\eta^3\pa_0\eta_3
+ e^{-i\mathcal{T}\tau}\eta^3\pa_1\te^- - e^{i\mathcal{T}\tau}
\eta_3\pa_1\te_- + \mathcal{T}\eta^3\eta_3 \nonumber \\
&+& im (e^{-i\mathcal{T}\tau}\eta^3\te^+ + e^{i\mathcal{T}\tau}\eta_3\te_+
+ e^{i\mathcal{T}\tau}\te^-\bar{\eta} + e^{-i\mathcal{T}\tau}\te_-\eta),
\label{pml}\end{eqnarray}
where $\te^+ = \te_+^{\dagger}, \te^- = \te_-^{\dagger}$ and the 
winding-number dependence appears, which is compared with the case A.
Under the shifts of $\te_- \rightarrow e^{-i\mathcal{T}\tau}\te_-,
\te_+ \rightarrow e^{i\mathcal{T}\tau}\te_+$ the expression (\ref{pml})
becomes
\begin{eqnarray}
\mathcal{L}_{2F} &=& i\te^+\pa_0\te_+ + i\bar{\eta}\pa_0\eta
- (\bar{\eta}\pa_1\te^+  - \eta \pa_1\te_+ ) - \mathcal{T}(\bar{\eta}\eta
+ \te^+\te_+) \nonumber \\
&+& i\te^-\pa_0\te_- + i\eta^3\pa_0\eta_3 + \eta^3\pa_1\te^- - 
\eta_3\pa_1\te_- + \mathcal{T}(\eta^3\eta_3 + \te^-\te_-) \nonumber \\
&+& im (e^{-2i\mathcal{T}\tau}\eta^3\te^+ 
+ e^{2i\mathcal{T}\tau}\eta_3\te_+ + e^{2i\mathcal{T}\tau}\te^-\bar{\eta}
+ e^{-2i\mathcal{T}\tau}\te_-\eta).
\label{ltm}\end{eqnarray}
The winding-number dependent terms have a large time-dependent phase in 
the large limit of the total angular momentum so that they oscillate and
average to zero as in \cite{AM,KT}. Therefore in the large $\mathcal{T}$
limit the $m$-dependent terms can be ignored. The resulting
Lagrangian shows a simple separated expression in the same way as 
(\ref{syl}) of the case A. 

Now for the case A we further set $\eta_-$ and $\te_3$ to zero in
(\ref{syl}) and (\ref{lfp}) in order to obtain a fermionic Lagrangian
with only two complex fermions $\eta_+, \te$
\begin{equation}
\mathcal{L}_F = i\eta^+\pa_0\eta_+ + i\bar{\te}\pa_0\te + 
\eta^+\pa_1\bar{\te} - \eta_+\pa_1\te - \mathcal{T}(\bar{\te}\te + 
\eta^+\eta_+). 
\label{lfa}\end{equation}
Introducing a two-component complex (Dirac) spinor $\psi$ by
combining the two complex fermions as 
\begin{equation}
\psi \equiv \left(\begin{array}{c} \psi_1 \\ \psi_2 \end{array} \right)
= \left(\begin{array}{c} \eta_+ \\ \bar{\te} \end{array} \right) \;
\mathrm{or} \; \left(\begin{array}{c} \te \\ \eta^+ \end{array} \right)
\end{equation}
we rewrite the fermionic Lagrangian (\ref{lfa}) as
\begin{equation}
\mathcal{L}_F = i(\psi_1^{\dagger}\pa_0\psi_1 + 
\psi_2^{\dagger}\pa_0\psi_2) + \psi_1^{\dagger}\pa_1\psi_2 - 
\psi_2^{\dagger}\pa_1\psi_1 - \mathcal{T}(\psi_1^{\dagger}\psi_1 -
\psi_2^{\dagger}\psi_2).
\end{equation}
Further it takes a Lorentz-invariant expression for a Dirac fermion with
mass $\mathcal{T}$ on the flat two-dimensional world-sheet 
\begin{equation}
\mathcal{L}_F = i\bar{\psi}\rho^{\mu}\pa_{\mu}\psi + \mathcal{T}
\bar{\psi}\psi 
\label{dir}\end{equation}
with $\rho^0 = -\sigma^3, \rho^1 = i\sigma^1$ and $\bar{\psi}= 
\psi^{\dagger}\rho^0$. This relativistic Lagrangian for the SU(1$|$2)
sector is compared with the non-relativistic quadratic fluctuation 
Lagrangian of one complex fermion for the 
SU(1$|$1) sector produced by integrating 
an extra fermion in ref. \cite{BST}.

In the fermionic action $S_F = \frac{\sqrt{\la}}{2\pi}\int d\tau d\sigma
 \mathcal{L}_F$ the overall factor $\sqrt{\la}$ can be removed by
the rescalings of 
\begin{equation}
\psi_{\alpha} \rightarrow \frac{\psi_{\alpha}}{\la^{1/4}}
\hspace{0.5cm} (\alpha = 1, 2).
\label{pss}\end{equation}
Moreover, the global $AdS_5$ time specified by $t = \mathcal{T}\tau$
in the large $\mathcal{T}$ limit yields 
\begin{equation}
S_F = \int dt \int \frac{d\sigma}{2\pi} \left[
i(\psi_1^{\dagger}\pa_0\psi_1 +\psi_2^{\dagger}\pa_0\psi_2) + 
\sqrt{\tilde{\la}}(\psi_1^{\dagger}\pa_1\psi_2 - 
\psi_2^{\dagger}\pa_1\psi_1) - (\psi_1^{\dagger}\psi_1 -
\psi_2^{\dagger}\psi_2) \right],
\label{sf}\end{equation}
where $\tilde{\la} = 1/\mathcal{T}^2 = \la/ J^2$ is the effective
BMN coupling constant. To create string states dual to the gauge theory
operators in the SU(1$|$2) sector we need to choose a proper 
representation of the anti-commutation relations for fermions.
The fermions are expanded in the Fourier modes 
\begin{equation}
\psi_{\alpha} = \sum_{n=-\infty}^{\infty} e^{in\sigma}\psi_{\alpha,n},
\hspace{1cm} \psi_{\alpha}^{\dagger} = \sum_{n=-\infty}^{\infty} 
e^{-in\sigma}\psi_{\alpha,n}^{\dagger}
\label{fmo}\end{equation}
by using the following creation and annihilation operators
\begin{equation}
\left(\begin{array}{c} \psi_{1n} \\ \psi_{2n} \end{array} \right) =
\left(\begin{array}{cc} f_n & g_n \\ g_n & f_n \end{array} \right)
\left(\begin{array}{c} a_n^- \\ b_n^+ \end{array} \right),
\hspace{1cm} 
\left(\begin{array}{c} \psi_{1n}^{\dagger} \\ \psi_{2n}^{\dagger} 
\end{array} \right) = \left(\begin{array}{cc} f_n & -g_n \\ -g_n & f_n
\end{array} \right) \left(\begin{array}{c} a_n^+ \\ b_n^- 
\end{array} \right),  
\label{cao}\end{equation}
where the functions $f_n, g_n$ are defined by
\begin{equation}
f_n = \sqrt{\frac{1}{2} + \frac{1}{2\omega_n} }, \hspace{1cm}
g_n = \frac{i\sqrt{\tilde{\la}}n}{1 + \omega_n}\sqrt{\frac{1}{2} + 
\frac{1}{2\omega_n} }
\label{fg}\end{equation}  
with $\omega_n = \sqrt{1 + \tilde{\la}n^2}$. The rotation matrices in
(\ref{cao}) also take  SU(2) forms. The substitution of (\ref{fmo})
into the action (\ref{sf}) yields
\begin{equation}
S_F = \int dt \sum_{n=-\infty}^{\infty}[ i(a_n^+\pa_0a_n^- 
+ b_n^+\pa_0b_n^- ) - \omega_n (a_n^+a_n^- + b_n^+b_n^-) ],
\label{pls}\end{equation}
which shows that $(a^-, a^+)$ and $(b^-, b^+)$ are pairs of canonically
conjugate fermionic operators and $\omega_n$ is the energy of a 
plane-wave state. The long SYM operators 
in the SU(1$|$2) sector are dual to
states obtained by acting operators $a_n^{\dagger}$ on the vacuum and
switching off the $b$ oscillators. Thus in the case of the non-point-like
circular string background we have produced the BMN-type plane-wave
Hamiltonian for the SU(1$|$2) sector, as a collection of free
massive fermionic oscillators.
The Lagrangian (\ref{pls}) shows the similar oscillator
expression to the quadratic plane-wave Lagrangian for the SU(1$|$1) sector
in ref. \cite{AAF}, where the superstring Hamiltonian with the near-plane
wave correction is constructed by using the uniform gauge and 
parametrizing the supercoset element in the different way from the one
used in \cite{MAT}. 

For the case B switching off $\te_-$ and $\eta_3$ for (\ref{ltm})
and (\ref{lft}) we have a reduced Lagrangian
\begin{equation}
\mathcal{L}_F = i\te^+\pa_0\te_+ + i\bar{\eta}\pa_0\eta - 
(\bar{\eta}\pa_1\te^+ - \eta\pa_1\te_+) - \mathcal{T}(\bar{\eta}\eta + 
\te^+\te_+).  
\end{equation}
In terms of the two complex fermions $\psi_1, \psi_2$ defined by
\begin{equation}
\psi \equiv \left(\begin{array}{c} \psi_1 \\ \psi_2 \end{array} \right)
= \left(\begin{array}{c} \te_+ \\ \bar{\eta} \end{array} \right) \;
\mathrm{or} \; \left(\begin{array}{c} \eta \\ \te^+ \end{array} \right)
\label{ren}\end{equation}
the Lagrangian is expressed as 
\begin{equation}
\mathcal{L}_F = i(\psi_1^{\dagger}\pa_0\psi_1 + 
\psi_2^{\dagger}\pa_0\psi_2) - \psi_1^{\dagger}\pa_1\psi_2 + 
\psi_2^{\dagger}\pa_1\psi_1 - \mathcal{T}(\psi_1^{\dagger}\psi_1 -
\psi_2^{\dagger}\psi_2).
\label{lfb}\end{equation}
If we rename the world-sheet coordinate $\sigma$ as $\sigma \rightarrow
-\sigma$ the Lagrangian again takes the same relativistic expression for
the Dirac fermion $\psi$ as (\ref{dir}). The fermions $\psi_\alpha, 
\psi_\alpha^{\dagger}$ are expanded in the Fourier modes in the same way
as (\ref{fmo}) and (\ref{cao}), where $f_n$ and $g_n$ are now given by
\begin{equation}
f_n = \sqrt{\frac{1}{2} + \frac{1}{2\omega_n} }, \hspace{1cm}
g_n = \frac{-i\sqrt{\tilde{\la}}n}{1 + \omega_n}\sqrt{\frac{1}{2} + 
\frac{1}{2\omega_n} }, \hspace{1cm}  \omega_n = \sqrt{1 + \tilde{\la}n^2},
\end{equation}  
which are obtained from (\ref{fg}) by $n \rightarrow -n$.
The substitution of this mode expansion into the action for (\ref{lfb})
 leads to the same plane-wave action as (\ref{pls}).

\section{Fermionic fluctuations over a circular string with two
unqual spins}

Let us consider a circular string background with two unequal spins 
$J_1, J_2$ and winding numbers $m_1, m_2$ \cite{ART}
\begin{equation}
X_i = a_i e^{i\omega_i \tau + im_i\sigma}, \hspace{1cm}
\omega_i^2 = m_i^2 + \nu^2, \hspace{1cm} \sum_{i=1}^{2}a_i^2 = 1,
\label{sou}\end{equation}
whose classical energy $\mathcal{E}$ and $\mathcal{T}_i =
\omega_i a_i^2 \;(i=1,2)$ are characterized by
\begin{equation}
\mathcal{E}^2 = 2\sum_{i=1}^{2} \omega_i\mathcal{T}_i - \nu^2,
\hspace{1cm} \sum_{i=1}^{2} m_i\mathcal{T}_i =0.
\label{en}\end{equation}
For the case A we substitute the bosonic background solution (\ref{sou}) 
into the fermionic Lagrangian (\ref{twl}) and (\ref{fol}) with
$\kappa/\sqrt{2}=1$ to have 
\begin{eqnarray}
\mathcal{L}_{2F} &=& i\sum_{i=1}^2 \eta^i\pa_0\eta_i + i\te^3\pa_0\te_3
+ i\bar{\te}\pa_0\te  \nonumber \\
&+& (a_1e^{-i\delta_1}\eta^2 - a_2e^{-i\delta_2}\eta^1)\pa_1\te^3 - 
(a_1e^{i\delta_1}\eta_2 - a_2e^{i\delta_2}\eta_1)\pa_1\te_3 \nonumber \\
&+& (a_1e^{i\delta_1}\eta^1 + a_2e^{i\delta_2}\eta^2)\pa_1\bar{\te} - 
(a_1e^{-i\delta_1}\eta_1 + a_2e^{-i\delta_2}\eta_2)\pa_1\te \nonumber \\ 
&+& (\omega_1a_1^2 - \omega_2a_2^2)(\eta_1\eta^1 - \eta_2\eta^2) 
\nonumber \\
&+& a_1a_2(\omega_1 + \omega_2)(e^{-i\delta_1 + i\delta_2}\eta_1\eta^2 + 
e^{i\delta_1 - i\delta_2}\eta_2\eta^1), \nonumber \\
\mathcal{L}_{4F} &=& -4\eta_1\eta^1\eta_2\eta^2, 
\label{unl}\end{eqnarray}
where $\delta_i = \omega_i\tau + m_i\sigma \; (i=1,2)$.

Performing the following SU(2)-type rotation of fermionic variables
suggested from the terms including the $\sigma$-derivative in (\ref{unl})
\begin{equation}
\left(\begin{array}{c} \eta_- \\ \eta_+ \end{array} \right) =
\left(\begin{array}{cc} a_2e^{i\delta_2} & -a_1e^{i\delta_1} 
\\ a_1e^{-i\delta_1} & a_2e^{-i\delta_2} \end{array} \right)
\left(\begin{array}{c} \eta_1 \\ \eta_2 \end{array} \right), 
\label{sur}\end{equation}
we rewrite the Lagrangian (\ref{unl}) as
\begin{eqnarray}
\mathcal{L}_{2F} &=& i\eta^+\pa_0\eta_+ + i\eta^-\pa_0\eta_- + 
i\te^3\pa_0\te_3 + i\bar{\te}\pa_0\te \nonumber \\
&-& \eta^-\pa_1\te^3 + \eta_-\pa_1\te_3 - \eta^+\pa_1\bar{\te} - 
\eta_+\pa_1\te + 2(\omega_1a_1^2 + \omega_2a_2^2)(\eta_+\eta^+ - 
\eta_-\eta^-) \nonumber \\
&+&  2a_1a_2(\omega_1 - \omega_2)(e^{i(\delta_1 + \delta_2)}\eta_+\eta^- +
 e^{-i(\delta_1 + \delta_2)}\eta_-\eta^+),  \nonumber \\
\mathcal{L}_{4F} &=& -4\eta^+\eta_+\eta^-\eta_-.
\end{eqnarray}
Under the redefinition the coupling terms have a phase and the mass
terms are arranged to have a suitable mass parameter $\omega_1a_1^2 + 
\omega_2a_2^2 = \mathcal{T}_1 + \mathcal{T}_2 = \mathcal{T}$.
The large angular momentum expansion for (\ref{sou}) gives 
$\nu^2 = \mathcal{T}^2 - \sum_{i=1}^2 m_i^2\mathcal{T}_i/\mathcal{T}
+ \cdots$ and
\begin{equation}
\omega_1 = \mathcal{T} + \frac{(m_1^2 - m_2^2)\mathcal{T}_2}
{2\mathcal{T}^2} + \cdots, \hspace{1cm}
\omega_2 = \mathcal{T} - \frac{(m_1^2 - m_2^2)\mathcal{T}_1}
{2\mathcal{T}^2} + \cdots.
\label{ome}\end{equation}
For the large total angular momentum $\mathcal{T}$ that means large
$\omega_i \;(i=1,2)$, the coupling terms with the time-dependent phase
can be ignored since they average to zero. Then we have a
simple diagonalized expression
\begin{eqnarray}
\mathcal{L}_{2F} &=& i\eta^+\pa_0\eta_+ + i\bar{\te}\pa_0\te + 
\eta^+\pa_1\bar{\te} - \eta_+\pa_1\te - 
2\mathcal{T}\eta^+\eta_+ \nonumber \\
&+& i\eta^-\pa_0\eta_- + i\te^3\pa_0\te_3 - 
\eta^-\pa_1\te^3 + \eta_-\pa_1\te_3 + 2\mathcal{T}\eta^-\eta_-,
\label{ssl}\end{eqnarray}
which is compared with (\ref{lpm}).
Here making the rescalings of fermions 
 $\eta_+ \rightarrow e^{-i\mathcal{T}\tau}\eta_+, \eta_- \rightarrow 
e^{i\mathcal{T}\tau}\eta_-$, we observe that $\mathcal{L}_{4F}$ is not
changed and $\mathcal{L}_{2F}$ is transformed into the same expression as
(\ref{lpm}). The rescalings combine with the $\tau$- and 
$\sigma$-dependent rotation (\ref{sur}) into an SU(2)-type rotation
\begin{equation}
\left(\begin{array}{c} \eta_- \\ \eta_+ \end{array} \right) =
\left(\begin{array}{cc} a_2e^{-i\mathcal{T}\tau + i\delta_2} & 
-a_1e^{-i\mathcal{T}\tau +i\delta_1} \\ a_1e^{i\mathcal{T}\tau -i\delta_1}
 & a_2e^{i\mathcal{T}\tau -i\delta_2} \end{array} \right)
\left(\begin{array}{c} \eta_1 \\ \eta_2 \end{array} \right), 
\end{equation}
which reduces to the previous $\sigma$-dependent rotation (\ref{cha})
for the equal spin case $\omega_1=\omega_2 = \mathcal{T}, m_1=-m_2=-m,
a_1=a_2= 1/\sqrt{2}$. Therefore the compensating  redefinitions of 
$\te \rightarrow
e^{i\mathcal{T}\tau}\te, \te_3 \rightarrow e^{-i\mathcal{T}\tau}\te_3$
for the transformed $\mathcal{L}_{2F}$ yield the same symmetric
separated expression as (\ref{syl}).
Thus for the case A reduced with $\eta_- = \te_3 =0$ in the circular
string background with two unequal spins the quantum plane-wave
spectrum (\ref{pls}) can be reproduced again through the suitable
renaming of the fermions. The fermionic string configuration with the
$a$ oscillator only is considered to correpond to the long SYM operator
which consists of the large and different number of two complex scalar 
fields and a few fermions in the SU(1$|$2) sector.

Now we turn our attention to the case B. The substitution of circular
string solution (\ref{sou}) into the  fermionic Lagrangian 
(\ref{twl}) and (\ref{fol}) with $\kappa/\sqrt{2}=1$ gives again
(\ref{lft}) for $\mathcal{L}_{4F}$ and an involved expression 
for $\mathcal{L}_{2F}$ which suggests the following particular 
choice of field redefinitions
\begin{equation}
\left(\begin{array}{c} \te_- \\ \te_+ \end{array} \right) =
\left(\begin{array}{cc} a_2e^{i\delta_2} & -a_1e^{i\delta_1} 
\\ a_1e^{-i\delta_1} & a_2e^{-i\delta_2} \end{array} \right)
\left(\begin{array}{c} \te_1 \\ \te_2 \end{array} \right), 
\label{thr}\end{equation}
which has the same transformation matrix as (\ref{sur}) amd
resembles (\ref{chb}).
The inversion of (\ref{thr}) leads to the quadratic Lagrangian
\begin{eqnarray}
\mathcal{L}_{2F} &=& i\te^+\pa_0\te_+ + i\bar{\eta}\pa_0\eta
- (\bar{\eta}\pa_1\te^+  - \eta \pa_1\te_+ ) - \mathcal{T}(\bar{\eta}\eta
+ \te^+\te_+) + im_a(\bar{\eta}\te^+ + \eta\te_+) \nonumber \\
&+& i\te^-\pa_0\te_- + i\eta^3\pa_0\eta_3 + \eta^3\pa_1\te^- - 
\eta_3\pa_1\te_- + \mathcal{T}(\eta^3\eta_3 + \te^-\te_-) +
im_a(\eta^3\te^- + \eta_3\te_-) \nonumber \\
&-& a_1a_2(\omega_1 - \omega_2)(e^{-i(\delta_1 + \delta_2)}\te^+\te_- +
 e^{i(\delta_1 + \delta_2)}\te^-\te_+) \nonumber \\
&+& ia_1a_2(m_1 - m_2)[e^{i(\delta_1 + \delta_2)}(\bar{\eta}\te^- -
\eta_3\te_+) + e^{-i(\delta_1 + \delta_2)}(\eta\te_- - \eta^3\te^+)]
\nonumber \\
&-& a_1a_2(\omega_1 - \omega_2)(e^{-i(\delta_1 + \delta_2)}\eta\eta^3
 +  e^{i(\delta_1 + \delta_2)}\eta_3\bar{\eta})
\label{lal}\end{eqnarray}
with $m_a = m_1a_1^2 + m_2a_2^2$ which is compared with 
$\mathcal{T} = \omega_1a_1^2 + \omega_2a_2^2$. It is confirmed that
this expression indeed reduces to (\ref{ltm}) for the equal spin case.
In this case the SU(2)-type rotation (\ref{thr}) can be expressed as
\begin{equation}
\left(\begin{array}{c} \te_-e^{-i\mathcal{T}\tau} \\ 
\te_+e^{-i\mathcal{T}\tau} \end{array} \right) =
\frac{1}{\sqrt{2}} \left(\begin{array}{cc} e^{im\sigma} & -e^{-im\sigma} 
\\ e^{im\sigma} & e^{-im\sigma}  \end{array} \right)
\left(\begin{array}{c} \te_1 \\ \te_2 \end{array} \right), 
\end{equation} 
which is just the product of the previous $\sigma$-dependent rotation
(\ref{chb}) and the succeeding shifts of $\te_- \rightarrow 
e^{-i\mathcal{T}\tau}\te_-, \te_+ \rightarrow e^{i\mathcal{T}\tau}\te_+$.
The quartic Lagrangian $\mathcal{L}_{4F}$ remains the same expression
as (\ref{lft}). Taking the fast-string limit of the quadratic Lagrangian
 $\mathcal{L}_{2F}$ (\ref{lal}) 
we see that the involved coupling terms with the exponential phase
factors $e^{\pm i(\delta_1 + \delta_2)}$ average to zero.
In the resulting  $\mathcal{L}_{2F}$ the fermionic system $(\te_+, \eta)$
is separated from the fermionic one  $(\te_-, \eta_3)$,
and each system has the non-zero winding-number dependent terms
with a coefficient $m_a$, which vanish for the equal spin case. 
 
Here putting $\te_- = \eta_3 =0$ for the reduction which keeps only two
complex fermions $\te_+, \eta$, we have
\begin{equation}
\mathcal{L}_{F} = i\te^+\pa_0\te_+ + i\bar{\eta}\pa_0\eta
- (\bar{\eta}\pa_1\te^+  - \eta \pa_1\te_+ ) - \mathcal{T}(\bar{\eta}\eta
+ \te^+\te_+) + im_a(\bar{\eta}\te^+ + \eta\te_+).
\end{equation}
Through the renaming of fermionic fields in the same way as (\ref{ren})
$\mathcal{L}_F$ becomes 
\begin{equation}
\mathcal{L}_F = i(\psi_1^{\dagger}\pa_0\psi_1 + 
\psi_2^{\dagger}\pa_0\psi_2) - \psi_1^{\dagger}(\pa_1 \pm im_a)\psi_2 + 
\psi_2^{\dagger}(\pa_1 \pm im_a)\psi_1 - \mathcal{T}
(\psi_1^{\dagger}\psi_1 - \psi_2^{\dagger}\psi_2), 
\label{mlf}\end{equation}
which is expressed through the renaming of $\sigma \rightarrow
- \sigma$ as
\begin{equation}
\mathcal{L}_F = i\bar{\psi}\rho^{\mu}\pa_{\mu}\psi + \mathcal{T}
\bar{\psi}\psi \pm m_a\bar{\psi}\rho^1\psi,
\end{equation}
where $+$ corresponds to the one choice $\psi = \left(\begin{array}{c} 
\te_+ \\ \bar{\eta} \end{array} \right)$ and $-$ to the other choice 
$\psi = \left(\begin{array}{c} \eta \\ \te^+ \end{array} \right)$,
and the bosonic background dependence is specified by $\mathcal{T}$
and $m_a$.   We substitute the mode expansion of $\psi_{\alpha}, 
\psi_{\alpha}^{\dagger}$, (\ref{fmo}) with (\ref{cao}) into the
fermionic action for (\ref{mlf}). 
If the following parametrization is chosen
\begin{equation}
f_n = \sqrt{\frac{1}{2} + \frac{1}{2\omega_n} }, \hspace{1cm}
g_n = \frac{-i\sqrt{\tilde{\la}}(n \pm m_a)}{1 + \omega_n}
\sqrt{\frac{1}{2} +  \frac{1}{2\omega_n} }, \hspace{1cm}  
\omega_n = \sqrt{1 + \tilde{\la}(n \pm m_a)^2},
\end{equation}   
the same plane-wave action as (\ref{pls}) is derived.
For the two unequal spin case we have observed that there seems a 
difference for the energy spectrum $\omega_n$ of each mode
between the reduced A system $(\eta_+, \te)$ and the reduced
B system $(\te_+, \eta)$. The $\omega_n$ for the reduced B system
is specified with the mode number $n$ shifted by the winding-number
dependent factor $m_a$. However, using (\ref{ome}) we estimate
$m_a$ in the $\tilde{\la} = 1/\mathcal{T}^2$ expansion as
\begin{equation}
m_a = \frac{m_1\mathcal{T}_1}{\omega_1} + \frac{m_2\mathcal{T}_2}
{\omega_2} = \frac{m_1\mathcal{T}_1 + m_2\mathcal{T}_2}{\mathcal{T}}
- \frac{1}{\mathcal{T}^2} \frac{(m_1 + m_2)(m_1 - m_2)^2\mathcal{T}_1
\mathcal{T}_2}{2\mathcal{T}^2} + \cdots,
\end{equation}
whose first leading term is zero through (\ref{en}).
To the leading order in $\tilde{\la}$, that is, in the plane-wave limit
both the reduced A system and the reduced B system show the same
energy spectrum as Fock-space states for the fermionic fluctuation
around the circular string background with two unequal spins.

For the ``SU(2$|$2)" string theory sector we make a system by combining
two cases A and B, that is, putting $X_3 = 0$ only in 
the fermionic Lagrangian $\mathcal{L}_{2F}$ (\ref{twl})  and 
$\mathcal{L}_{4F}$ (\ref{fol}), and take the 
large total angular momentum limit 
for the sum of the quadratic fermionic Lagrangians (\ref{ssl})
and (\ref{lal}). The quartic Lagrangian becomes $\mathcal{L}_{4F}
= \eta^+\eta_+\bar{\eta}\eta$ when the reduction specified by
both $\eta_- = \te_3 =0$ and $\eta_3 = \te_- =0$ is taken.
However, owing to the rescalings of fermionic fields by $1/\la^{1/4}$
such as (\ref{pss}) the quartic action becomes of the order 
$1/\sqrt{\la}$. In the leading large $\sqrt{\la}$ approximation
near the classical bosonic solution with two spins we need
to know only the quadratic part.
The reduced quadratic action is shown to give the plane-wave spectra
for the two fermions in the SU(2$|$2) sector by switching off the
two relevant $b$ oscillators.

\section{Conclusion}

From the SU(3)$\times$U(1) invariant superstring action in $AdS_5
\times S^5$ which is produced by choosing the conformal gauge for the
$\kappa$-symmetry gauge fixed superstring action, we have constructed
the truncated action for the SU(1$|$2) sector which describes the 
fermionic fluctuations over a circular string background with 
two angular momenta and two winding numbers.

The suitable SU(2)-type $\tau$- and $\sigma$-dependent rotation of
fermionic fields and the first-string limit simplify the starting 
superstring action such that the involved coupling terms can be
ignored and there remain two separated massive fermion systems.
The mass terms have been characterized by the total angular momentum
that arises from an adequate combination of the coupling terms
between the fermionic fluctuations and the bosonic background fields.
We have observed that the appropriate renamings of fermionic fields lead
to a Lorentz-invariant action for a massive 2d Dirac fermion and its
plane-wave spectrum for the SU(1$|$2) sector. For the truncated systems
A and B for the two unequal spin case there appears a difference in
the plane-wave spectra, but the difference specified by the winding
numbers can be neglected to the leading order in $\tilde{\la}$.
Combining the two truncated systems we have also constructed 
the leading plane-wave action for the SU(2$|$2) sector.

Recently in \cite{FPZ} in the uniform light-cone gauge the bosonic
and fermionic quantum fluctuation spectra have been constructed in the
near plane-wave limit for the SU(1$|$2) and SU(2$|$3) sectors,
whose string configuration with one large angular momentum and no
winding numbers is dual to the long SYM operator with a few bosonic
impurity  $W$ fields and a few fermionic fields in a large number of 
bosonic $Z$ fields, while our string 
configuration with two large angular momenta
and two winding numbers for the SU(1$|$2) and SU(2$|$2) sectors
is dual to the long SYM operator with a large number of bosonic
impurity $W$ fields and a few fermionic fields in a large number of 
bosonic $Z$ fields. In spite of such differences for the bosonic 
backgrounds and the gauge choices, both fermionic leading spectra 
show the same BMN-type behavior in the limit 
of the large angular momenta.


\begin{thebibliography}{99}
\bibitem{MGW} J.M. Maldacena, ``The large N limit of superconformal
field theories and supergravity,'' Adv. Theor. Math. Phys. \textbf{2}
(1998) 231 [hep-th/9711200]; S.S. Gubser, I.R. Klebanov and A.M. Polyakov,
``Gauge theory correlators from non-critical string theory,"
Phys. Lett. \textbf{B428} (1998) 105 [hep-th/9802109]; E. Witten, 
``Anti-de Sitter space and holography,"
Adv. Theor. Math. Phys. \textbf{2} (1998) 253 [hep-th/9802150].
\bibitem{MT} R.R. Metsaev, ``Type IIB Green-Schwarz superstring in
plane wave Ramond-Ramond background," Nucl. Phys. \textbf{B625}
(2002) 70 [hep-th/0112044];  
R.R. Metsaev and A.A. Tseytlin, ``Exactly solvable 
model of superstring in plane wave Ramond-Ramond background," Phys.
Rev. \textbf{D65} (2002) 126004 [hep-th/0202109].
\bibitem{BMN} D. Berenstein, J.M. Maldacena and H. Nastase, 
``Strings in flat space and pp waves from $\mathcal{N}$=4 super
Yang Mills," JHEP \textbf{04} (2002) 013 [hep-th/0202021].
\bibitem{GKP} S.S. Gubser, I.R. Klebanov and A.M. Polyakov,
``A semi-classical limit of the gauge/string correspondence,"
Nucl. Phys. \textbf{B636} (2002) 99 [hep-th/0204051].
\bibitem{FT} S. Frolov and A.A. Tseytlin, ``Semiclassical 
quantization of rotating superstring in $AdS_5\times S^5$," JHEP
\textbf{06} (2002) 007 [hep-th/0204226]; ``Multi-spin string solutions
in $AdS_5 \times S^5$," Nucl. Phys. \textbf{B668} (2003) 77 
[hep-th/0304255]; ``Quantizing three-spin
string solution in $AdS_5 \times S^5$," JHEP \textbf{07} (2003) 016
[hep-th/0306130]; ``Rotating string 
solutions: AdS/CFT duality in non-supersymmetric sectors," Phys. Lett. 
\textbf{B570} (2003) 96 [hep-th/0306143].
\bibitem{BMS} N. Beisert, J.A. Minahan, M. Staudacher and K. Zarembo, 
``Stringing spins and spinning strings," JHEP \textbf{09}
(2003) 010 [hep-th/0306139];
N. Beisert, S. Frolov, M. Staudacher and A.A. Tseytlin, 
``Precision spectroscopy of AdS/CFT," JHEP \textbf{10} (2003) 037 
[hep-th/0308117];
G. Arutyunov, S. Frolov, J. Russo and A.A. Tseytlin, 
``Spinning strings in $AdS_5 \times S^5$ and integrable systems,"
Nucl. Phys. \textbf{B671} (2003) 3 [hep-th/0307191].
\bibitem{ART} G. Arutyunov, J. Russo and A.A. Tseytlin, 
``Spinning strings in $AdS_5\times S^5$: new integrable system relations,"
Phys. Rev. \textbf{D69} (2004) 086009 [hep-th/0311004].
\bibitem{AT} A.A. Tseytlin, ``Spinning strings and AdS/CFT duality,"
 hep-th/0311139.
\bibitem{MZ} J.A. Minahan and K. Zarembo, ``The Bethe-ansatz for 
$\mathcal{N} =4$ super Yang-Mills," JHEP \textbf{03} (2003) 013 
[hep-th/0212208].
\bibitem{NBS} N. Beisert, ``The complete one-loop dilatation operator
of $\mathcal{N} =4$ super Yang-Mills theory," Nucl. Phys. \textbf{B676}
(2004) 3 [hep-th/0307015]; N. Beisert and M. Staudacher, ``The 
$\mathcal{N}=4$ SYM integrable super spin chain," Nucl. Phys. 
\textbf{B670} (2003) 439 [hep-th/0307042].
\bibitem{NB} N Beisert, ``The su$(2|3)$ dynamic spin chain," Nucl. Phys.
\textbf{B682} (2004) 487 [hep-th/0310252].
\bibitem{BZ} N. Beisert, ``The dilatation operator of 
$\mathcal{N} = 4$ super Yang-Mills theory and integrability," 
Phys. Rept. \textbf{405} (2005) 1 [hep-th/0407277];
K. Zarembo, ``Semiclassical Bethe ansatz and AdS/CFT," Comptes Rendus
Physique \textbf{5} (2004) 1081, Fortsch, Phys. \textbf{53}
(2005) 647 [hep-th/0411191];
J. Plefka, ``Spinning strings and integrable spin chains in the
AdS/CFT correspondence, hep-th/0507136.
\bibitem{KMM} V.A. Kazakov, A. Marshakov, J.A. Minahan and K. Zarembo, 
``Classical/quantum integrability in AdS/CFT," JHEP \textbf{05}
(2004) 024 [hep-th/0402207].
\bibitem{KZ} V.A. Kazakov and K. Zarembo, 
``Classical/quantum integrability in non-compact sector of AdS/CFT," 
JHEP \textbf{10} (2004) 060 [hep-th/0410105];
N. Beisert, V.A. Kazakov and K. Sakai, ``Algebraic curve for the SO(6)
sector of AdS/CFT," hep-th/0410253;
S. Sch\"afer-Nameki, ``The algebraic curve of 1-loop planar 
$\mathcal{N} = 4$ SYM," Nucl. Phys. \textbf{714} (2005) 
3 [hep-th/0412254];
N. Beisert, V.A. Kazakov, K. Sakai and K. Zarembo,
``The algebraic curve of classical superstrings on $AdS_5 \times S^5$,"
hep-th/0502226.
\bibitem{BDS} N. Beisert, V. Dippel and M. Staudacher, ``A novel long
range spin chain and planar $\mathcal{N} =4$ super Yang-Mills,"
JHEP \textbf{07} (2004) 075 [hep-th/0405001].
\bibitem{AFS} G. Arutyunov, S. Frolov and M. Staudacher, ``Bethe
ansatz for quantum strings," JHEP \textbf{10} (2004)
016 [hep-th/0406256].
\bibitem{CLM} C.G. Callan, H.K. Lee, T. McLoughlin, J.H. Schwarz, 
I. Swanson and X. Wu, ``Quantizing string theory in $AdS_5 \times S^5$:
Beyond the pp-wave," Nucl. Phys. \textbf{B673} (2003) 3 [hep-th/0307032];
C.G. Callan, T. McLoughlin and I. Swanson, ``Holography beyond the 
Penrose limit," Nucl. Phys. \textbf{B694} (2004) 115 [hep-th/0404007]; 
``Higher impurity AdS/CFT correspondence in the near-BMN limit," Nucl. 
Phys. \textbf{B700} (2004) 271 [hep-th/0405153].
\bibitem{RSS} A. Rej, D. Serban and M. Staudacher ``Planar $\mathcal{N}=4$
gauge theory and the Hubbard model," hep-th/0512077.
\bibitem{MS} M. Staudacher, ``The factorized S-matrix of CFT/AdS,"
JHEP \textbf{05} (2005) 054 [hep-th/0412188].
\bibitem{BS} N. Beisert and M. Staudacher, ``Long-range PSU(2,2$|$4) Bethe
ansatze for gauge theory and strings," Nucl. Phys. \textbf{B727} (2005) 1
[hep-th/0504190].
\bibitem{CHM} C.G. Callan, J. Heckman, T. McLoughlin and I. Swanson,
``Lattice super Yang-Mills: A virial approach to operator dimensions,"
Nucl. Phys. \textbf{B701} (2004) 180 [hep-th/0407096].
\bibitem{NBT} N. Beisert, ``The su(2$|$2) dynamic S-matrix," 
hep-th/0511082; ``An su(1$|$1)-invariant S-matrix with dynamic 
representations," hep-th/0511013;
R.A. Janik, ``The $AdS_5 \times S^5$ superstring worldsheet
S-matrix and crossing symmetry," hep-th/0603038.
\bibitem{BIZ} B.I. Zwiebel, ``$\mathcal{N}=4$ SYM to two loops: Compact
expressions for the non-compact symmetry algebra of the su(1,1$|$2) 
sector," hep-th/0511109.
\bibitem{FPT} S.A. Frolov, I.Y. Park and A.A. Tseytlin, ``On one-loop 
correction to energy of spinning strings in $S^5$," Phys. Rev. 
\textbf{D71} (2005) 026006 [hep-th/0408187];
I.Y. Park A. Tirziu and A.A. Tseytlin, ``Spinning strings in
$AdS_5 \times S^5$: One-loop correction to energy in SL(2) sector,"
JHEP \textbf{03} (2005) 013 [hep-th/0501203].
\bibitem{SZ} S. Sch\"afer-Nameki, M. Zamaklar and K. Zarembo, ``Quantum
corrections to spinning strings in $AdS_5 \times S^5$ and Bethe ansatz:
A comparative study," JHEP \textbf{09} (2005) 051 [hep-th/0507189]; 
S. Sch\"afer-Nameki and M. Zamaklar, ``Stringy sums and corrections
to the quantum string Bethe ansatz," JHEP \textbf{10} (2005) 044 
[hep-th/0509096]; 
N. Beisert and A.A. Tseytlin, ``On quantum corrections to spinning 
strings and Bethe equations," Phys. Lett. \textbf{B629} (2005) 102 
[hep-th/0509084];
S. Sch\"afer-Nameki, ``Exact expressions for quantum corrections to 
spinning strings," hep-th/0602214.
\bibitem{AM} A. Mikhailov, ``Speeding strings," JHEP \textbf{12} (2004)
058 [hep-th/0311019]; ``Slow evolution of nearly-degenerate extremal 
surfaces," J. Geom. Phys. \textbf{54} (2005) 228 hep-th/0402067; 
``Supersymmetric null-surfaces," JHEP \textbf{09} (2004) 068 
[hep-th/0404173]; ``Notes on fast moving strings," hep-th/0409040.
\bibitem{MK} M. Kruczenski, ``Spin chains and string theory,"
Phys. Rev. Lett. \textbf{93} (2004) 161602 [hep-th/0311203]. 
\bibitem{KT} M. Kruczenski and A.A. Tseytlin, ``Semiclassical
relativistic strings in $S^5$ and long coherent operators in 
$\mathcal{N}=4$ SYM theory," JHEP \textbf{09} (2004) 038 
[hep-th/0406189].
\bibitem{KRT} M. Kruczenski, A.V. Ryzhov and A.A. Tseytlin, ``Large
spin limit of $AdS_5 \times S^5$ string theory and low energy expansion
of ferromagnetic spin chains," Nucl. Phys. \textbf{B692} (2004) 3
[hep-th/0403120];
R. Hernandez and E. Lopez, ``The SU(3) spin chain sigma model and 
string theory," JHEP \textbf{04} (2004) 052 [hep-th/0403139];
C. Kristjansen and T. Mansson, ``The circular, elliptic
three-spin string from the SU(3) spin chain," Phys. Lett. \textbf{B596}
(2004) 265 [hep-th/0406176].
\bibitem{ST} B. Stefanski, J. and A.A. Tseytlin, ``Large spin limits
of AdS/CFT and generalized Landau-Lifshitz equations," 
JHEP \textbf{05} (2004) 042 [hep-th/0404133].
\bibitem{BM} S. Bellucci, P.-Y. Casteill, J.F. Morales, C. Sochichiu,
``SL(2) spin chain and spinning strings on $AdS_5 \times S^5$,"
Nucl. Phys. \textbf{B707} (2005) 303 [hep-th/0409086];
S. Ryang ``Circular and folded multi-spin strings in spin chain
sigma models," JHEP \textbf{10} (2004) 059 [hep-th/0409217].
\bibitem{MTT} J.A. Minahan, A. Tirziu and A.A. Tseytlin, ``$1/J$ 
corrections to semiclassical AdS/CFT states from quantum Landau-Lifshitz
model," Nucl. Phys. \textbf{B735} (2006) 127 [hep-th/0509071];
``$1/J^2$ corrections to BMN energies from the quantum long range
Landau-Lifshitz model," JHEP \textbf{11} (2005) 031 [hep-th/0510080];
A. Tirziu, ``Quantum Landau-Lifshitz model at four loops:
$1/J$ and $1/J^2$ corrections to BMN energies," hep-th/0601139.
\bibitem{HL} R. Hernandez and E. Lopez, ``Spin chain sigma models with 
fermions," JHEP \textbf{11} (2004) 079 [hep-th/0410022]
\bibitem{BCM} S. Bellucci, P.-Y. Casteill and J.F. Morales, ``Superstring
sigma models from spin chains: the SU(1,1$|$1) case," Nucl. Phys.
\textbf{B729} (2005) 163 [hep-th/0503159].
\bibitem{BC} S. Bellucci and P.-Y. Casteill, ``Sigma model from
SU(1,1$|$2) spin chain," hep-th/0602007.
\bibitem{BST} B. Stefanski, J. and A.A. Tseytlin, ``Super spin chain 
coherent state actions and $AdS_5 \times S^5$ superstring," 
Nucl. Phys. \textbf{B718} (2005) 83 [hep-th/0503185].
\bibitem{RMT} R.R. Metsaev and A.A. Tseytlin, ``Type IIB superstring
action in $AdS_5 \times S^5$ background," Nucl. Phys. \textbf{B533} (1998)
109 [hep-th/9805028].
\bibitem{MAT} R.R. Metsaev and A.A. Tseytlin, ``Superstring action in
$AdS_5 \times S^5$: $\kappa$-symmetry light cone gauge,"
Phys. Rev. \textbf{D63} (2001) 046002 [hep-th/0007036];
R.R. Metsaev, C.B. Thorn and A.A. Tseytlin, ``Light-cone superstring
in AdS space-time," Nucl. Phys. \textbf{B596} (2001)
151 [hep-th/0009171].
\bibitem{RS} R. Roiban and W. Siegel, ``Superstrings on $AdS_5 \times 
S^5$ supertwistor space," JHEP \textbf{11} (2000) 024 [hep-th/0010104];
I. Bena, J. Polchinski and R. Roiban, ``Hidden symmetries
of the $AdS_5 \times S^5$ superstring," Phys. Rev. \textbf{D69} 
(2004) 046002 [hep-th/0305116];
A. Das, J. Maharana, A. Melikyan and M. Sato, ``The algebra of transition
matrices for the $AdS_5 \times S^5$ superstring," JHEP \textbf{12} (2004)
055 [hep-th/0411200];
L.F. Alday, G. Arutyunov and A.A. Tseytlin, ``On 
integrability of classical superstrings in $AdS_5 \times S^5$," 
JHEP \textbf{07} (2005) 002 [hep-th/0502240];
A. Das, A. Melikyan and M. Sato, ``The algebra of flat currents
for the string on $AdS_5 \times S^5$ in the light-cone gauge," 
JHEP \textbf{11} (2005) 015 [hep-th/0508183].
\bibitem{AAF} L.F. Alday, G. Arutyunov and S. Frolov, ``New 
integrable system of 2dim fermions from strings on 
$AdS_5 \times S^5$," JHEP \textbf{01} (2006) 078 [hep-th/0508140].
\bibitem{GAF} G. Arutyunov and S. Frolov, ``Uniform light-cone gauge
for strings in $AdS_5 \times S^5$: Solving su(1$|$1) sector,"
JHEP \textbf{01} (2006) 055 [hep-th/0510208].
\bibitem{TMS} T. McLoughlin and I. Swanson, ``N-impurity superstring
spectra near the pp-wave limit," Nucl. Phys. \textbf{B702} (2004) 86
[hep-th/0407240].
\bibitem{KKZ} T. Klose and K. Zarembo, ``Bethe ansatz in stringy
sigma models," hep-th/0603039.
\bibitem{FPZ} S. Frolov, J. Plefka and M. Zamaklar, ``The $AdS_5 
\times S^5$ superstring in light-cone gauge and its Bethe equation,"
hep-th/0603008.


\end{thebibliography}
\end{document}